\documentclass[12pt]{article}
\usepackage[super,comma,numbers,sort&compress]{natbib}  
\usepackage{flexisym}
                               
\usepackage{multicol}

\makeatletter \renewcommand\@biblabel[1]{#1.} \makeatother

\usepackage{tikz}
\usepackage{xcolor}

\usepackage{titlesec} 

\usepackage{graphicx}

\usepackage[english]{babel}
\addto\captionsenglish{}

\titleformat{\subsection}[runin]
{\normalfont\large\bfseries}{\thesubsection}{1em}{}
\titleformat{\subsubsection}[runin]
{\normalfont\normalsize\bfseries}{\thesubsubsection}{1em}{}

\newenvironment{Nabstract}{%
	\begin{quote} \bf}
	{\end{quote}}

\title{Experimental and Theoretical Realization of Zenneck Wave-based Non-Radiative, Non-Coupled Wireless Power Transmission }

\author
{Sai Kiran Oruganti,$^{1}$ Jagannath Malik,$^{1}$  Jongwon Lee,$^{1}$  Woojin Park,$^{1}$\\ Bonyoung Lee,$^{1}$ Seoktae Seo,$^{1}$ Dipra Paul,$^{1}$ Haksun Kim,$^{1}$ \\Thomas Thundat $^{2}$ and Franklin Bien.$^{1\ast}$ \\
	\\
	\normalsize{$^{1}$Ulsan National Institute of Science and Technology, Republic of Korea}\\
	\normalsize{$^{2}$University at Buffalo, New York, USA}\\
	\\
	\normalsize{$^\ast$To whom correspondence should be addressed; E-mail:  bien@unist.ac.kr.}
}

\date{}

\begin{document} 
	
	
	\baselineskip24pt
	
	
	\maketitle

	
	\begin{Nabstract}
%
	
A decade ago, non-radiative wireless power transmission re-emerged as a promising alternative to deliver electrical power to devices where a physical wiring proved to be unfeasible. However, existing approaches are neither scalable nor efficient when multiple devices are involved, as they are restricted by factors like coupling and external environments.  Zenneck waves are excited at interfaces, like surface plasmons and have the potential to deliver electrical power to devices placed on a conducting surface. Here, we demonstrate, efficient and long range delivery of electrical power by exciting non-radiative waves over metal surfaces to multiple loads. Our modeling and simulation using Maxwell’s equation with proper boundary conditions shows Zenneck type behavior for the excited waves and are in excellent agreement with experimental results. In conclusion, we physically realize a radically different power transfer system, based on a wave, whose existence has been fiercely debated for over a century.

	\end{Nabstract}
	
	
	In 2007, coupled WPT re-emerged as an alternative to deliver electrical power to systems where physical wiring is difficult or dangerous\cite{4,Kurs}. Since, then a number of notable articles appeared \cite{NC18,NC218,NC14}. However, these were improvements or at the best variations of the coupled WPT systems originally proposed in \cite{Kurs}. All these existing systems rely on critical coupling between coils of the transmitter and the receiver for efficient delivery of power. The resonance conditions are easily affected by the external factors\cite{01}. It has been also well understood that the need for a critical coupling leads to peak splitting phenomena for multiple resonant devices\cite{Ab}. This causes efficiency degradation and hence, are unsuitable for emerging fields such as, internet of things(IoT) and dynamic charging of electrical vehicles. It is therefore, Assawaworrarit et al, proposed parity time circuits to resolve the issue of dynamic wireless charging\cite{01}. Unfortunately, we will continue to face these limitations as long as we rely on critical coupling.\\
	
	A non-radiating \textit{wave-based} wireless power transfer system would be a desirable candidate to solve these issues. We wish to draw the attention to Zenneck wave(Sommerfeld-Zenneck wave), which resides at the metal-air interface, akin to SP and surface waves(SW)\cite{1,2}.
	While SP and surface wave (SW) have been widely researched areas in optical physics and metasurfaces, they are relatively less studied in the microwave regime\cite{Th,NY,JL,St}.  Likewise, much research around ZW is focused on the communications and geophysics applications\cite{Jan,JZG,Sam}. Unfortunately, ZW has been surrounded by the controversies regarding their physical existence\cite{Sc,Sar1,Sar2}. The bulk of the controversy arose from the alleged “sign error” committed by Sommerfeld in 1909\cite{Sar1,Sar2}. Some authors have shown feasibility of such waves by recreating the critical Seneca lake experiment to debunk the Sommerfeld sign error myth\cite{Cor}. Quite literally, one does not find any study on the utilization of ZW for non-radiative power transfer.
	Recently in 2014 and 2017 Sarkar et al, have taken great pains to clarify the confusions arising due to the definitions of SW, SP and ZW through their mathematically rigorous articles\cite{Sar1,Sar2}. The properties exhibited by ZW’s are like SW and SP, with certain differences. All these three physical phenomena are transverse magnetic (TM) modes and exhibit evanescent field decay away from the metal-air or metal-dielectric or conductive-dielectric interface. Unlike SW, the ZW come into existence as a result of zero TM reflection coefficient. SP come into existence at the quasi-particle levels. Whereas, ZW propagate in the form of localized charge oscillations. Just like SW and SP, when ZW are excited on metal surfaces, the net flow of current is zero.  The Brewster angle of incidence in case of ZW is frequency independent. Therefore, the attenuation of ZN waves is frequency independent and the attenuation rate is slow in the transverse direction\cite{Sar1,Sar2}. The equi-phases of ZN waves tilt backwards in the air (at the metal-air interface)\cite{1,2}. They sink into a lossy dielectric media, as mathematically demonstrated by Barlow and Cullens in their classic article\cite{Bar}. This sinking phenomenon was later experimentally demonstrated by Jangal et al and Ling et al\cite{Jan,Ling}.
	Here we demonstrate the physical realization of a ZW non-radiative power transmission using the arrangement of a planar ground backed impedance (GBI) surface and a half wave helical transformer at radio frequency (RF). The GBI structure establishes a TM wave spread across the metal surface, whereas, the half wave helical transformer drives the voltage across the GBI terminals. The helical transformer is like the telsa transformer (supplementary information). However, unlike the tesla transformer it does not generate standing waves. It was earlier theorized that an infinite vertical aperture is needed to excite a Zenneck wave and hence was not physically realizable\cite{hill}. In our results we demonstrate that the problem of infinite vertical aperture can be certainly bypassed. We also demonstrate that unlike the coupled non-radiative wireless power transmission systems, the presence of leaky metal shields does not affect the power transmission efficiency \cite{Kurs,Li}.
	Moreover, we demonstrate uniform power delivery to multiple receiving units with meaningful efficiency by theory and experiment, as we eliminate the frequency peak splitting issue altogether\cite{Ab}.

	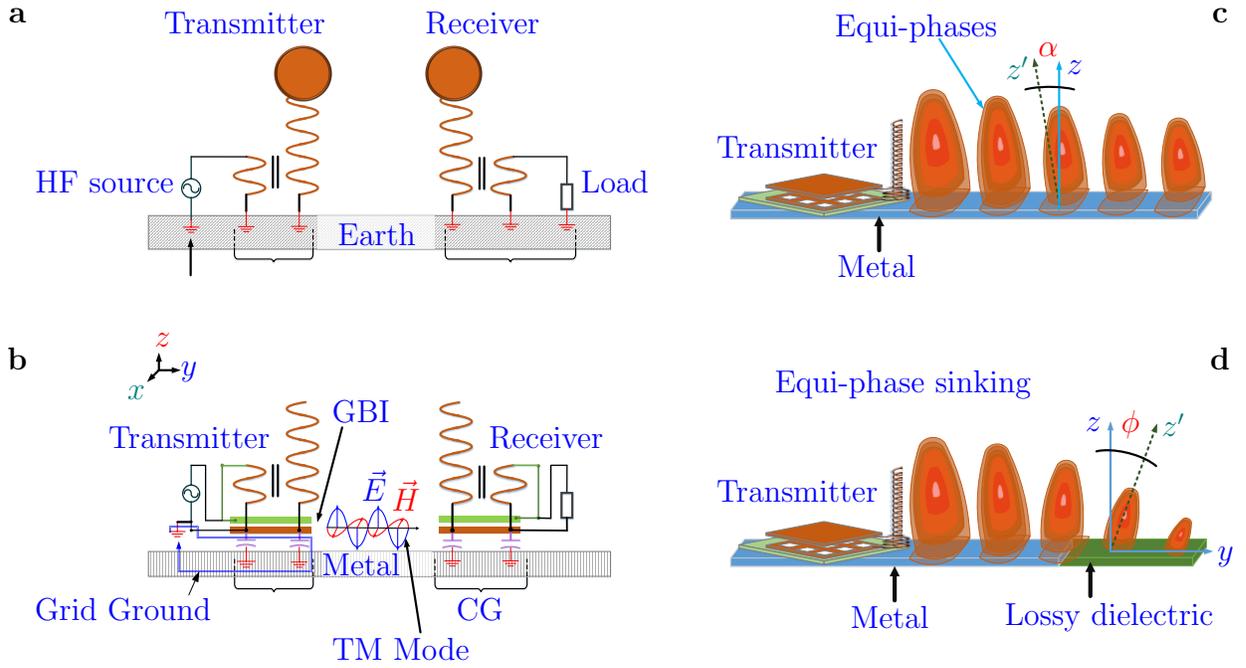
\begin{figure*}[]
		\centering
		\begin{tikzpicture}
		\node[above right] (img) at (0,0) {\includegraphics[width=16cm]{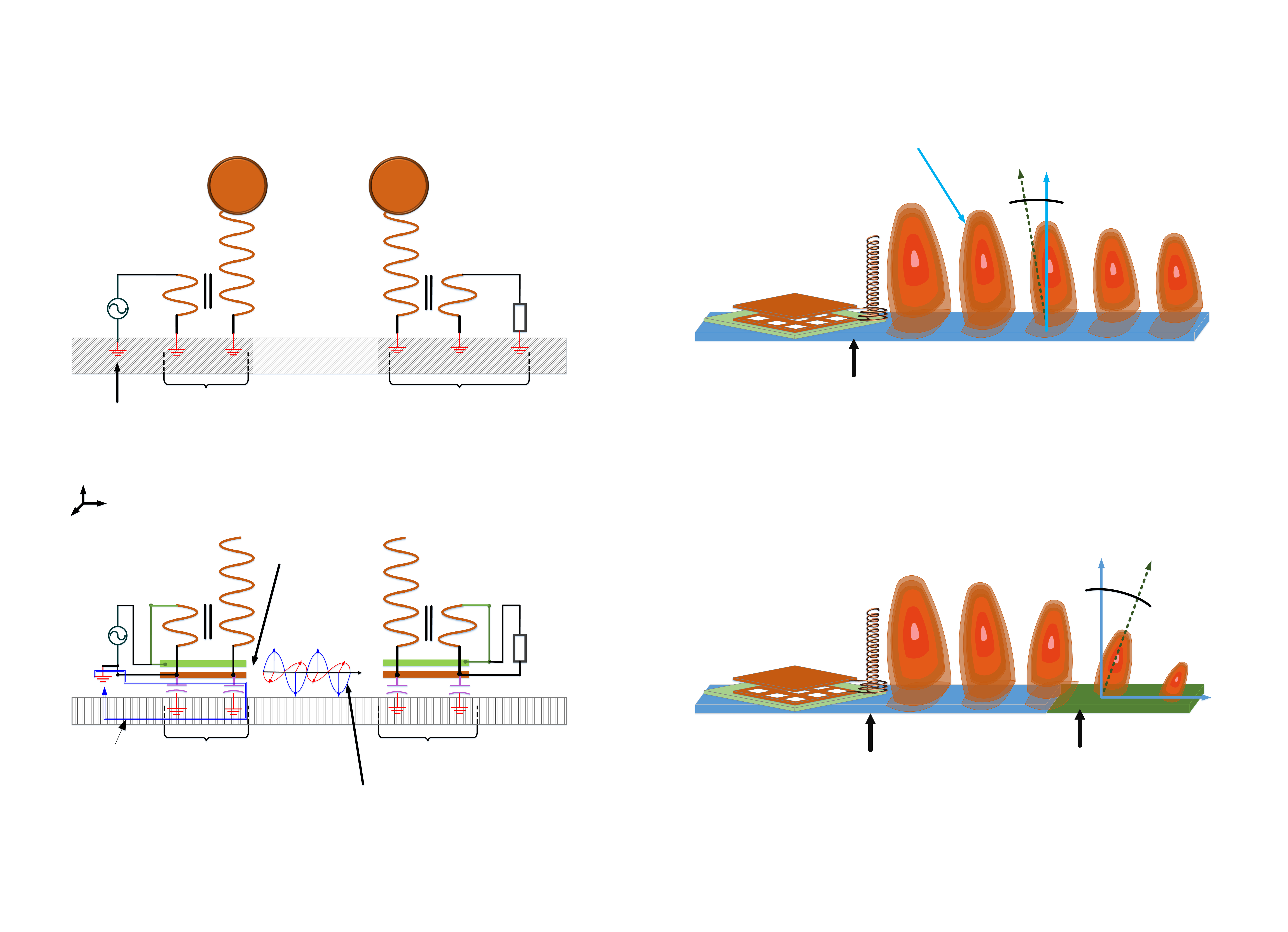}};



		
		\node at (-20pt, 290pt) {\textbf{a}};
		\node at (116pt,207pt) {{\color{blue}Earth}}; \node at (66pt,287pt) {\color{blue}Transmitter}; \node at (156pt,287pt) {\color{blue}Receiver};  \node at (13pt,227pt) {\color{blue}HF source};\node at (206pt,227pt) {\color{blue}Load};
\node at (-20pt, 160pt) {\textbf{b}};
\node at (35pt, 167pt) {{\color{red}$z$}}; \node at (45pt, 155pt) {{\color{blue}$y$}}; \node at (25pt, 147pt) {{\color{teal}$x$}};
		\node at (45pt, 130pt) {{\color{blue}Transmitter}};	\node at (180pt, 130pt){{\color{blue}Receiver}};
		
			\node at (20pt, 65pt) {{\color{blue}Grid Ground}};\node at (155pt, 65pt) {{\color{blue}CG}};
			\node at (125pt, 50pt) {{\color{blue}TM Mode}}; \node at (110pt, 82pt) {{\color{blue}Metal}};
			\node at (115pt, 112pt) {{\color{blue}$\vec{E}$}}; \node at (127pt, 108pt) {{\color{red}$\vec{H}$}}; \node at (110pt, 140pt) {{\color{blue}GBI}};
	\node at (435pt, 290pt) {\textbf{c}};
	\node at (320pt, 285pt) {{\color{blue}Equi-phases}}; \node at (370pt, 277pt) {{\color{red}$\alpha$}};
	\node at (275pt, 240pt) {{\color{blue}Transmitter}}; \node at (305pt, 195pt) {{\color{blue}Metal}};
	\node at (358pt, 270pt) {{\color{teal}$z\textprime$}};\node at (380pt, 270pt) {{\color{blue}$z$}};
\node at (435pt, 160pt) {\textbf{d}};
	\node at (401pt, 135pt) {{\color{red}$\phi$}}; \node at (393pt, 62pt) {{\color{blue}Lossy dielectric}};
	\node at (310pt, 62pt) {{\color{blue}Metal}};		\node at (275pt, 110pt) {{\color{blue}Transmitter}};
		\node at (315pt, 150pt) {{\color{blue}Equi-phase sinking}}; 	\node at (437pt, 85pt) {{\color{blue}$y$}}; \node at (417pt, 135pt) {{\color{teal}$z\textprime$}}; \node at (387pt, 135pt) {{\color{blue}$z$}};
		\end{tikzpicture}
		\vspace{-1.3cm}	\caption{ Concept of the proposed Zenneck wave system (a) In Tesla transformer, grounding is an extremely critical factor. Both the transmitter and receiver are grounded to the earth ground. (b) Approach followed in this study based on half wave helical transformer. The GBI resonator sets up a TM-Mode wave. Capacitive grounding[CG] is done through the metal, which in turn pulls the reference potential of the metal to the grid ground.(c)Localized field oscillation forms the Equi-phases, with an angle of backward tilt $\alpha$. In case of imperfect ground, this tilt is forward (d) The equi-phases undergo a "forward" tilt ($\phi$) in a lossy dielectric and sink.} 
		\label{Con}
	\end{figure*}

\begin{figure*}[]
	\centering
	\begin{tikzpicture}
	\node[above right] (img) at (0,0) {\includegraphics[width=15.5cm]{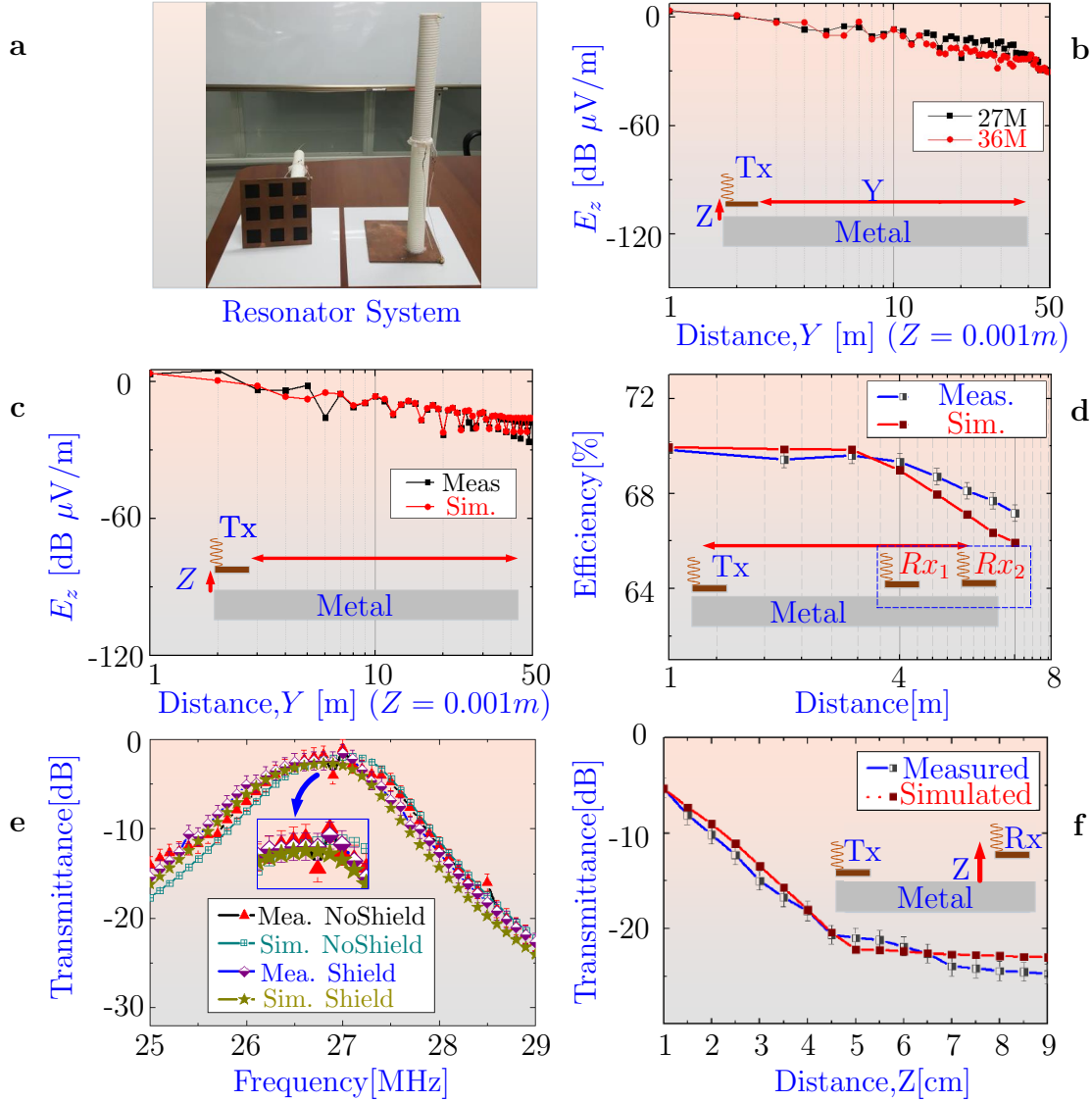}};

	
	\node at (0pt,410pt) {\textbf{a}};
		\node at (125pt,307pt) {\color{blue}Resonator System};
		
	
	\node at (410pt,410pt) {\textbf{b}};
		\node at (220pt,375pt) [rotate=90] {\color{blue} $E_{z}$ [dB $\mu$V/m]};
	\node at (245pt,422pt)  {0}; \node at (240pt,380pt)  {-60}; \node at (238pt,337pt)  {-120};
	
	\node at (252pt,310pt)  {1}; 	\node at (340pt,310pt)  {10}; \node at (398pt,310pt)  {50};
	\node at (330pt,298pt) {\color{blue}Distance,$Y$ [m] ($Z=0.001m$)};
	\node at (330pt,339pt) {\color{blue}Metal}; \node at (265pt,345pt) {\color{blue}Z};
	\node at (330pt,355pt) {\color{blue}Y}; \node at (380pt,383pt) {\color{black}\footnotesize{27M}};
	\node at (380pt,375pt) {\color{red}\footnotesize{36M}}; \node at (284pt,364pt) {\color{blue}Tx};
	\node at (410pt,270pt) {\textbf{d}};	
	\node at (220pt,230pt) [rotate=90]{\color{blue}Efficiency[$\%$]};
	\node at (240pt,275pt) {72}; \node at (240pt,237pt) {68}; \node at (240pt,200pt) {64};
	\node at (252pt,167pt) {1}; \node at (340pt,167pt) {4}; \node at (400pt,167pt) {8};
	\node at (330pt,156pt) {\color{blue}Distance[m]};
	\node at (370pt,276pt) {\color{blue}Meas.};	\node at (370pt,265pt) {\color{red}Sim.};
		\node at (275pt,210pt) {\color{blue}Tx};\node at (350pt,210pt) {\color{red}$Rx_1$}; \node at (379pt,210pt) {\color{red}$Rx_2$}; \node at (305pt,192pt) {\color{blue}Metal};
	\node at (410pt,110pt) {\textbf{f}};
		\node at (220pt,88pt) [rotate=90]{\color{blue}Transmittance[dB]};
	\node at (240pt,145pt) {0}; \node at (236pt,105pt) {-10}; \node at (236pt,70pt) {-20};
	\node at (330pt,10pt) {\color{blue}Distance,Z[cm]}; \node at (250pt,25pt) {1}; \node at (268pt,25pt) {2}; 
	\node at (287pt,25pt) {3};\node at (305pt,25pt) {4}; \node at (324pt,25pt) {5};\node at (342pt,25pt) {6};
	\node at (361pt,25pt) {7};\node at (380pt,25pt) {8}; \node at (398pt,25pt) {9};
	\node at (366pt,132pt) {\color{blue}Measured};\node at (366pt,122pt) {\color{red}Simulated}; 
	\node at (354pt,82pt) {\color{blue}Metal}; \node at (364pt,93pt) {\color{blue}Z}; \node at (326pt,100pt) {\color{blue}Tx}; \node at (388pt,105pt) {\color{blue}Rx};
	\node at (0pt,270pt) {\textbf{c}};	
	\node at (41pt,280pt) {0}; 	
	\node at (40pt,229pt) {-60};
	\node at (38pt,175pt) {-120};	
	\node at (85pt,225pt) {\color{blue}Tx}; 
	\node at (53pt,168pt) {1}; 	
	\node at (139pt,168pt) {10};
	\node at (200pt,168pt) {50};
	\node at (18pt,225pt) [rotate=90]{\color{blue} $E_{z}$ [dB $\mu$V/m]};

	\node at (65pt,205pt) {\color{blue}$Z$}; \node at (130pt,195pt) {\color{blue}Metal};
	\node at (130pt,156pt) {\color{blue}Distance,$Y$ [m] ($Z=0.001m$)};
	\node at (85pt,225pt) {\color{blue}Tx}; \node at (175pt,242pt) {\color{black}\footnotesize{Meas}}; \node at (175pt,233pt) {\color{red}\footnotesize{Sim.}};

	\node at (0pt,110pt) {\textbf{e}};
		\node at (18pt,90pt) [rotate=90]{\color{blue}Transmittance[dB]};	\node at (45pt,142pt) {0}; \node at (41pt,108pt) {-10};\node at (41pt,74pt) {-20}; \node at (41pt,40pt) {-30};
	\node at (51pt,25pt) {25};  \node at (88pt,25pt) {26};  \node at (125pt,25pt) {27}; \node at (164pt,25pt) {28}; \node at (200pt,25pt) {29}; \node at (125pt,10pt) {\color{blue}Frequency[MHz]};
	\node at (125pt,75pt) {\footnotesize{Mea. NoShield}}; \node at (125pt,64pt) {\color{teal}\footnotesize{Sim. NoShield}};\node at (120pt,53pt) {\color{blue}\footnotesize{Mea. Shield}};
	\node at (120pt,43pt) {\color{olive}\footnotesize{Sim. Shield}};

	\end{tikzpicture}
	\caption{ Experiment and Simulation results: Zenneck Wave at metal-air interface (a) Ground Backed Impedance resonator system, with a half wavelength helical coil.  (b) Experimental results of the Z component of the Electric field in the Y-direction 1 to 50 $m$, shows a slow attenuation rate. Two resonators with frequencies 27 and 36 $MHz$ were designed and compared. The resonators were placed at a vertical distance of $Z=0.001m$ above the metal surface.(c) Measured and simulated results comparison of E field attenuation along Y-direction at 27 $MHz$; $Z=0.001m$.  (d) Multi receiver power transfer efficiency. (e) Experimental and simulated results of the transmittance parameters, when transmitting and receiving unit are under shield and no-shield conditions. (f) Evanescent field decay experiment. }
	\label{NC2}
\end{figure*}

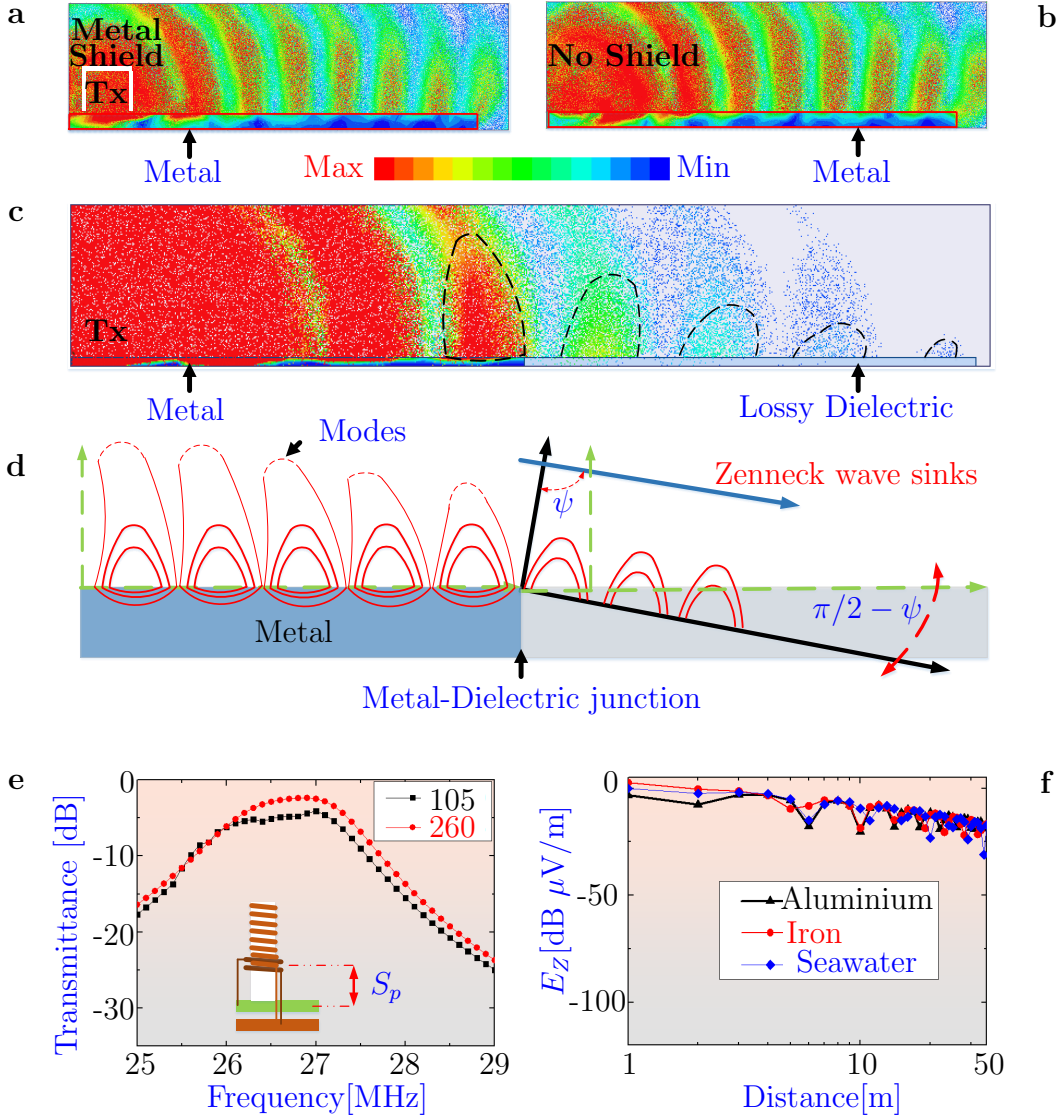
\begin{figure*}[]
	\centering
	\begin{tikzpicture}
	\node[above right] (img) at (0,0) {\includegraphics[width=12.7cm]{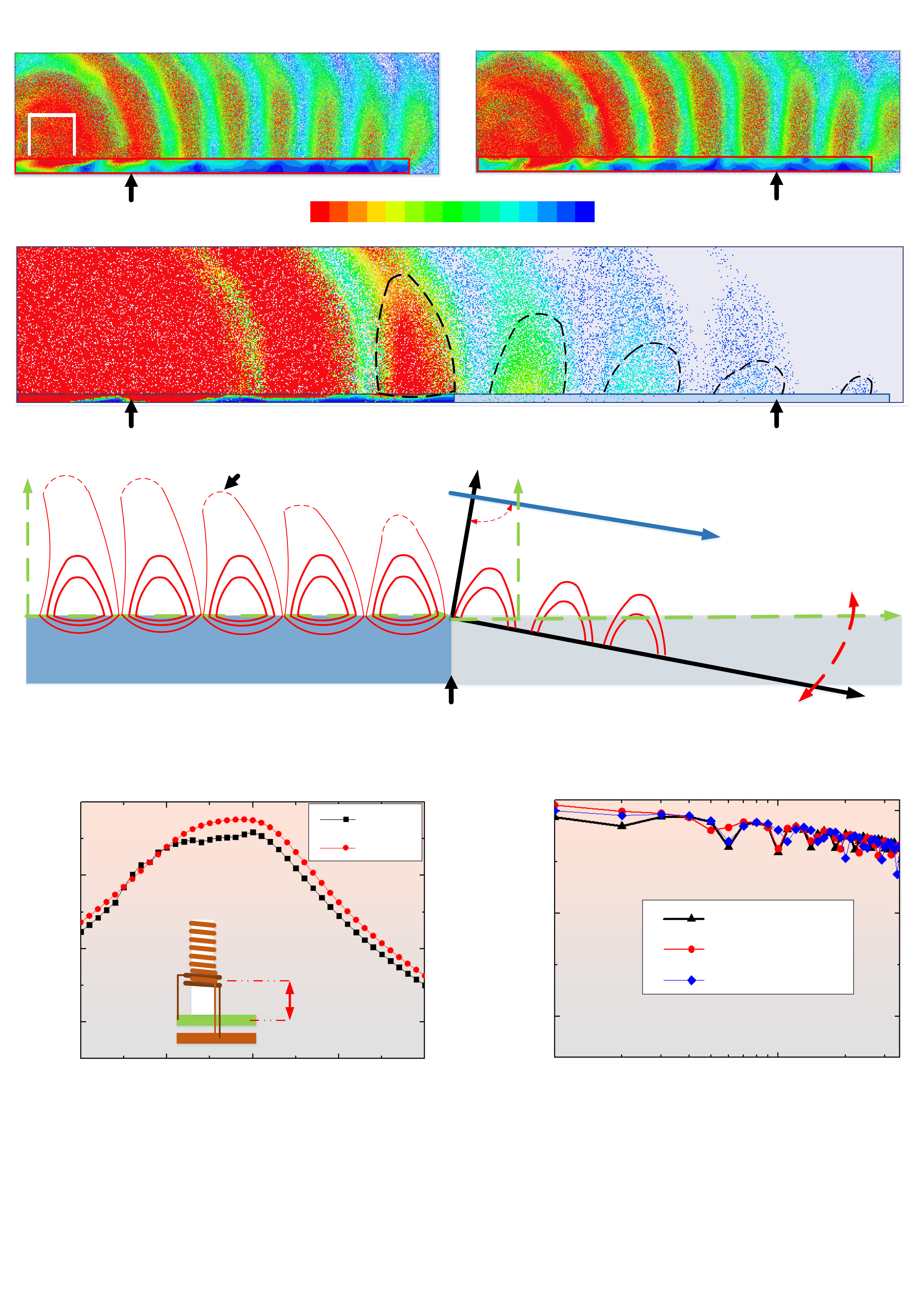}};
	
	\node at (-10pt,495pt)  {\textbf{a}};
	\node at (28pt,490pt)  {\textbf{\color{black}Metal}}; 
	\node at (28pt,480pt)  {\textbf{Shield}};
	\node at (24pt,466pt)  {\textbf{Tx}};
	\node at (53pt,436pt)  {\color{blue}Metal};
	\node at (110pt,438pt)  {\color{red}Max};
	\node at (250pt,438pt)  {\color{blue}Min};
	
	\node at (380pt,495pt)  {\textbf{b}};
	\node at (220pt,480pt)  {\textbf{No Shield}};
	\node at (306pt,436pt)  {\color{blue}Metal};
	\node at (-10pt,420pt)  {\textbf{c}};
	\node at (304pt,346pt)  {\color{blue}Lossy Dielectric};
	\node at (54pt,346pt)  {\color{blue}Metal};
	\node at (24pt,376pt)  {\textbf{Tx}};
	
	\node at (-10pt,325pt)  {\textbf{d}};
	\node at (304pt,322pt)  {\color{red}Zenneck wave sinks};
	\node at (95pt,262pt)  {\color{black}Metal};
	\node at (185pt,236pt)  {\color{blue}Metal-Dielectric junction};
	\node at (121pt,338pt)  {\color{blue}Modes};
	\node at (197pt,310pt)  { \color{blue}$\psi$};
	\node at (312pt,270pt)  { \color{blue}$\pi/2-\psi$};
	\node at (-10pt,205pt)  {\textbf{e}};
	\node at (155pt,198pt) {\color{black}105};                               
	\node at (155pt,188pt) {\color{red}260};                               
	\node at (37pt,98pt) {25};    
	\node at (70pt,98pt) {26};  
	\node at (103pt,98pt) {27};
	\node at (137pt,98pt) {28};
	\node at (171pt,98pt) {29};            
	\node at (130pt,127pt) {\color{blue} $S_{p}$};
	\node at (10pt,151pt) [rotate=90] {\color{blue}Transmittance [dB]};   
	\node at (27pt,119pt)  {-30};   
	\node at (27pt,147pt)  {-20};   
	\node at (27pt,177pt)  {-10};  
	\node at (32pt,205pt)  {0};    
	\node at (104pt,85pt)  {\color{blue}Frequency[MHz]};   
	\node at (380pt,205pt)  {\textbf{f}};
	\node at (194pt,160pt) [rotate=90] {\color{blue}$E_Z$[dB $\mu$V/m]};   
	\node at (215pt,205pt) {0};   
	\node at (210pt,162pt) {-50};   
	\node at (210pt,122pt) {-100};
	
	\node at (222pt,98pt) {1};    
	\node at (310pt,98pt) {10}; 
	\node at (358pt,98pt) {50};    
	\node at (308pt,161pt) {Aluminium}; 
	\node at (292pt,148pt) {\color{red}Iron};  
	\node at (308pt,137pt) {\color{blue}Seawater}; 
	
	\node at (295pt,85pt)  {\color{blue}Distance[m]};                                                           
	\end{tikzpicture}
	\vspace{-3 cm}\caption{ FEM Simulation Model. (A) E-field mode excitation, when transmitter(Tx) housed inside a partial metal shield. (B) E-field mode excitation when there is no shield. The Tx is electrically isolated from the metal at all times. (C) E-field mode excitation when the first half is metal and the other half is a lossy dielectric with $\epsilon_{r}$=4. (D) Representative diagram of the mode profiles, completing in free space. The Zenneck wave sinks into the lossy dielectric. (E) Transmittance parameters when the coil is placed at a vertical spacer of $S_{p}=$105 $mm$ and 260 $mm$ from the ground layer of the resonator. (F) Attenuation of E-field in the transverse direction along the interface of Air-Aluminium, Air-Iron, Air-Seawater. }
	\label{NC3}
\end{figure*}


\begin{figure*}[]
	\centering
	\begin{tikzpicture}
	\node[above right] (img) at (0,0) {\includegraphics[width=15.5cm]{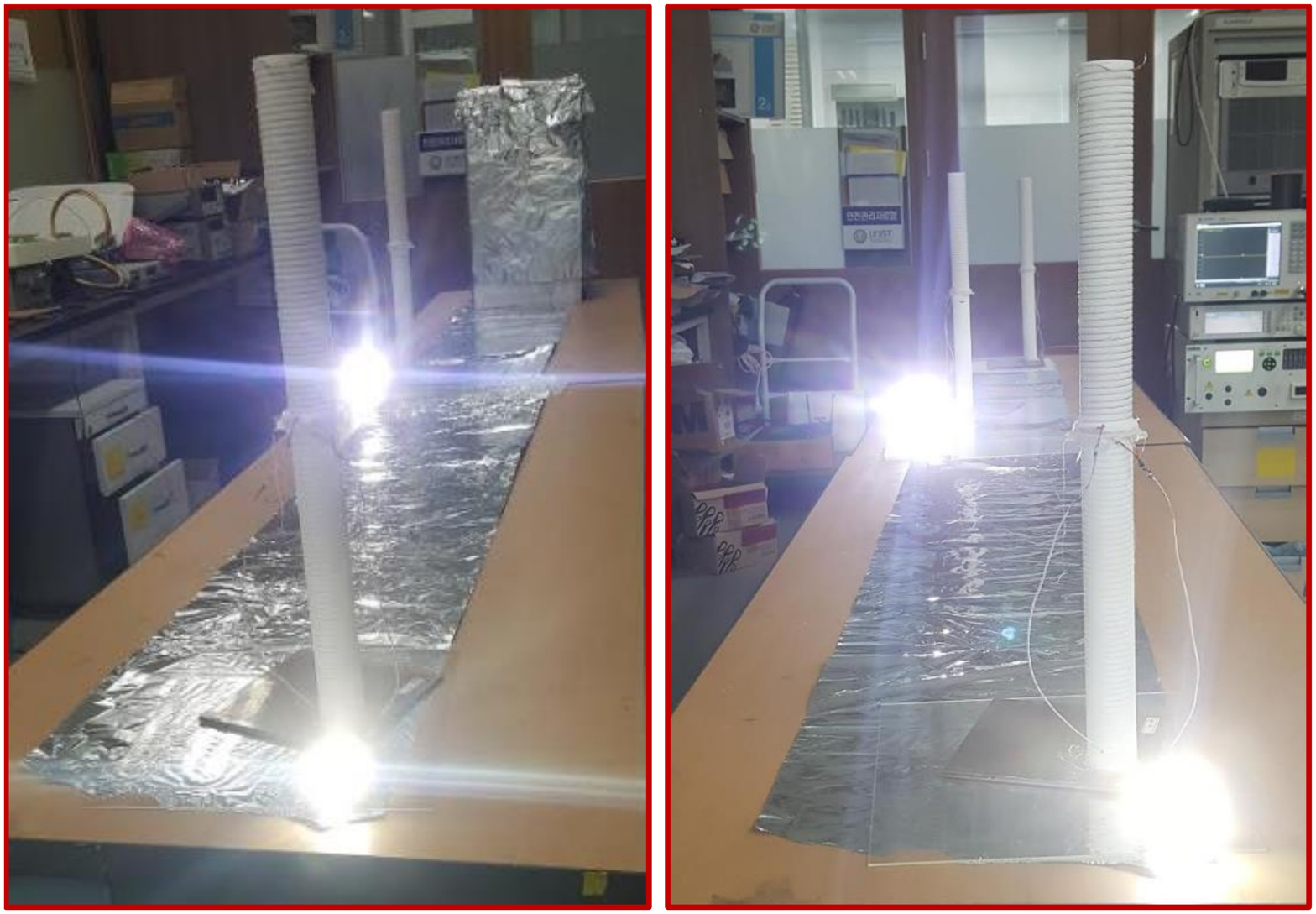}};
	
	\node at (117pt,5pt) {(a)}; \node at (325pt,5pt) {(b)};
	\end{tikzpicture}
	\caption{Power Transmission to multiple receivers under (a) shielded (b) Open conditions. Overall load of 40 $watts$}
	\label{PTO}
\end{figure*}

	While efficient transmission of non-radiative, wireless power over long distances using earth as a conductor is far from practical realization, it may be possible to utilize already existing metal structures to send guided mode waves for powering electrical devices\cite{4,5,6}. There exist many practical scenarios consisting of metallic infrastructures, such as nuclear plants, railway tracks, space ships, steel building structures, pipelines, etc. Practical applications include powering Internet of things (IoT) devices, charging for -marine vessels, smart manufacturing floors, and secured shipping containers\cite{5,6,19}. 

\section*{Results} 
Please note, that the experimental setup is described in the section 1 of the supplementary material. \\
The key concept of this study is presented in Fig.\ref{Con}, the detailed analytical model and solution is presented in the supplementary material(under section: analytical formalism). The Fig. \ref{Con} a, shows the mechanism of Tesla transformer based wireless power transfer system. The primary coil consists of low number of turns, while the secondary has large number of turns(quarter-wave). One end of the secondary is left freely suspended in the air. Sometimes, a toroid is attached to the free end of the secondary to restrict the electric field buildup to prevent discharges. The primary and the secondary coils on both the transmission and the receiving end share the common ground.  The generator, which operates as a high frequency AC source, is also grounded to the grid, which is in-turn grounded to the earth.\\ 
\subsubsection*{Approach followed in this study.} The Fig. \ref{Con} b  shows the schematic diagram of the system to excite zenneck waves at metal-air interface. Apart from exciting TM- waves using the GBI resonators at the metal-air interface, we use two critical concepts of Tesla transformer, namely-  half wave helical coil (Tesla transformer uses a quarter wave coil), in the secondary to build high potential differences across the terminals of the resonator and grounding the coils to the grid ground, capacitively via the metal. This pulls the reference potential of the metal to the same level as the grid ground. Thus, metal is transformed into a neutral entity\cite{19}. The metal's behavior as a neutral point is covered in the supplementary material(section 3). In the supplementary material(section 4) the analytical model is treated as a combination of vertical and horizontal hertzian dipoles(equations S7 to S27). The equation S26 and S27 are the expressions for the ZW field profiles in terms of Hertzian potential vector.
\subsubsection*{Equi-phases.} The Fig. \ref{Con} c, shows the equi-phases generated due to the localized field oscillations on the metal-air interface. The phase velocity of the wave in the metal is faster than the free space, hence a backward tilt with an angle $\alpha$ is observed, in accordance with \cite{1}.
 \subsubsection*{Sinking of Equi-phases.} Likewise, in the Fig. \ref{Con} d, the equi-phases undergo a forward tilt and subsequent sinking when they encounter a lossy dielectric\cite{Jan,Bar,Ling}.
\subsection*{Hallmark of Zenneck waves.} The ZW properties of the proposed system have been experimentally observed and are presented in Fig.\ref{NC2}. The resonator system is shown in the Fig. \ref{NC2} a,  dimensions and parameters are presented in the supplementary material(Fig S 10 and table ST 2).
\subsubsection*{a. Frequency independent slow attenuation rate.} The Fig. \ref{NC2} b, shows the comparison of the attenuation rate of the E-field[$dB \mu V/m$] for the two resonator systems designed for operating frequencies of 27 $MHz$ and 36 $MHz$. The transmitter and receiver was fixed at a height of $Z=0.001m$ above the metal surface. However, the receiver was moved along the interface ($Y$-direction) and the corresponding values were recorded\cite{Sar1}.  It is observed that the E-field values along the metal show a slow attenuation rate, independent of the frequency. This property is consistent with the ZW's as reported by Schelkunoff, Sarkar et al. and Barlow \cite{Sc,Sar1,Sar2,Bar}.
The Fig \ref{NC2} c, shows the measured and simulated results of the attenuation rate at 27 $MHz$. The simulation was done using Ansys high-frequency structure simulator(HFSS). It is observed that the experiments and simulation model are in excellent agreement.
\subsubsection*{b. Multi receiver efficiency.} The fig \ref{NC2} d, shows the multi receiver efficiency from 1 $m$ to 8 $m$ distance. Two receivers with identical loads were used of 20 watts each. The simulation result of the transmittance parameters are listed in the supplementary material (section 5). It is observed that the system efficiency varies between 66$\%$ to 62 $\%$ for a range of 1 to 8 $m$. The power transfer metrics at 8 m and 15 m are listed in the supplementary material(tables ST 6 and ST 7)
\subsubsection*{c. Leaky or partial metal shields.} The Fig. \ref{NC2} e, shows the comparison of measured and simulated results of the transmittance parameters under leaky shielded and non-shielded conditions. The transmittance parameters were observed using the state-of-the-art vector network analyzer. The FEM model is in good agreement with the measured results. It is observed that the proposed system, unlike the coupled WPT systems, has the ability to perform without any significant efficiency degradation\cite{Kurs, Li}.\\
\subsubsection*{d. Evanescent field/exponential decay.} An exponential E-field decay is also observed in the normal direction away from the metal-air interface(listed in Fig. \ref{NC2} f), consistent with the evanescent property of the ZW's\cite{1,2,Th,Jan,Sc,Sar1,Sar2,Cor,Bar,Ling}.

\subsubsection*{e. FEM Simulation model} The equi-phases mentioned in the Fig. \ref{Con} c, is shown in the Fig. \ref{NC3} a and b. These simulations were carried out for shielded and open field conditions. In line with the transmittance parameter results, the shield does not have a significant effect on the equi-phases.
\subsubsection*{f. Sinking of Equi-phases: Simulation.} The simulation of the sinking of the equi-phases of a zenneck wave in to a lossy dielectric was mentioned in Fig \ref{Con} d. This particular case is simulated using our FEM model and is shown in Fig. \ref{NC3} c, the simulation conditions are shown in Fig. \ref{NC3} d.
\subsubsection*{g. Eddy current effect.} The current carried in the primary of the resonator coil, is effected by the eddy currents generated on the metal. This effect was reduced by increasing the spacing between the coil and the GBI resonator from 105 $mm$ to 260 $mm$.
\subsubsection*{h. Attenuation rate along different interfaces.}  If the proposed method is exciting Zenneck waves at the metal-air interface, then, they must also show similar properties across various other conductive media. The Fig. \ref{NC3} f, shows the attenuation characteristics across aluminium (conductivity, $\sigma=3.8 \times 10^7 S/m$), iron($\sigma=1.03 \times 10^7 S/m$) and sea-water($\sigma=4 S/m$ and $\epsilon =81$). It is observed that the attenuation rate in seawater is faster than metal.

\subsection*{Power Transfer Demonstration.} The Fig.\ref{PTO} shows 40 $watts$ power transmission to multiple receiver units under shielded and open conditions across 2.7 $m$ of metal surface. The power transfer data and metrics are provided in the supplementary material in details. The supplementary material lists the metrics of transmission to distances of 8 and 15 meters.

\section*{Discussions}
We have demonstrated the excitation of waves on metal surfaces that can be used for delivering electrical power to multiple devices. The waves show slow transverse attenuation property similar to ZW's along the metal-air interface and can be used for efficient powering of devices up to 8 meters using this arrangement. We also show that ZW's can be used for transmitting power across partial metal enclosures. Therefore, the resonator system has the ability to overcome electromagnetic shielding and can be used for powering devices under leaky metal enclosures since the excited waves are non-radiative in nature. Power transmission to multiple receiving resonators with uniform efficiency has also been established experimentally and shows excellent agreement with simulation. The simulation result is compared with coupled mode power transfer system in the supplementary(Fig. S5). Existing coupled mode WPT systems undergo peak splitting when multiple receiving units are involved. Our study shows that using a wave-based mode of transmission, we can solve this issue.\\
 The efficiency of power transmission increases when multiple receiving units are present, as the power is uniformly spread across the metal surface. This kind of increase, due to multiple receiving units was also observed in a widely followed article, where weakly coupled WPT system is used \cite{Kurs2}. The maximum value of E and H-field emitted by the system is 34 $\%$ and 89 $\%$ lower than the permitted values, regulated by the ICNIRP guidelines at this frequency(Supplementary material Fig. S 11, S 12 and Table ST 3). Thus, this system should not pose as an occupational hazard to human operators. The proposed system has no effect on other devices operating in vicinity(supplementary material demo video links).

\end{document}